\documentclass[proceedings]{JHEP}
\usepackage{latexsym}
\usepackage{epsf}
\usepackage{amssymb}
\def\be{\begin{equation}}
\def\ee{\end{equation}}
\def\bea{\begin{eqnarray}}
\def\eea{\end{eqnarray}}

\def\pd{\partial}

\def\b{\beta}

\def\m{\mu}
\def\n{\nu}

\conference{Nonpertubative Quantum Effects 2000}

\title{Holography and the C-Theorem}

\author{Enrique \'Alvarez and C\'esar G\'omez\\
        Instituto de F{\'\i}sica Te\'orica \\
        Universidad Aut\'onoma de Madrid \\
        C-XVI, C.U. Cantoblanco \\
        28049 Madrid, Spain \\
        E-mail: \email{Enrique.Alvarez@uam.es},  
                \email{Cesar.Gomez@uam.es}}

\abstract{We review the geometric definition of C-function in the context
of field theories that admit a holographic gravity dual. 
}

\begin{document}
\section{Holographic C-Theorem}
\subsection{The Holographic Hypothesis}
The holographic hypothesis \cite{thooftsusskind} postulates that the number of
different quantum states in a given volume $V$ can not exceed
$e^{\frac{A}{4G_{D}}}$ where $A$ is the area of the corresponding boundary
$\partial V $
and $G_{D}$ the Newton constant in $D$ space-time dimensions. This 
hypothesis is
based on the idea that the maximun entropy associated with a volume $V$
is given by the Bekenstein-Hawking entropy $S=\frac{A}{4 G_{D}}$. 
Following 't Hooft 
we can define the number of holographic degrees of freedom associated
with a volume $V$ as  
\be
N_{hdof}=\frac{A}{4 ln2 G_{D}}
\ee
On the other hand, Maldacena's conjecture \cite{malda,witten,gubser} 
stablishes a 
correspondence
between $N=4$ supersymmetric Yang Mills in flat Minkowski space-time, with
gauge group $SU(N)$, and type IIB supergravity on $AdS_{5}\times S^{5}$ provided both
the $AdS_{5}$ and the $S^5$ radii are given by
\be
R=l_{s}(g_{s}N)^{\frac{1}{4}}
\ee
In order to check the holographic hypothesis for Maldacena's correspondence,
we will start (following Susskind and Witten, \cite{susskindwitten}) 
by considering $N=4$ supersymmetric Yang Mills in a box of
topology $S^{3}\times \mathbb{R}$, where $\mathbb{R}$ represents time and 
the radius of the sphere $S^{3}$ is taken to be equal one. 
Let us now introduce an \textbf{ultraviolet}
cutoff of size $\delta$. The number of degrees of freedom of
the regularized theory would then be given by:
\be
N^{YM}= \frac{N^{3}}{{\delta}^{3}}
\ee
Next we move into the $AdS_{5}$ gravity dual. The metric of
$AdS_{5}$ can be written as follows:
\be
ds^{2} = \frac{R^{2}}{(1-r^{2})^{2}}(4 dX^{2} -(1+r^{2})^{2} dt^{2})
\ee
In these coordinates the boundary is located at $r=1$. Notice that
the holographic coordinate is now represented by $r$. We can introduce an
\textbf{infrared} regulator by putting 
the regularized boundary at $r=1-\delta$.
The area of the regularized boundary is given by
\be
A(\delta)= \frac{R^{3}}{\delta^{3}}
\ee
Using the holographic hypothesis the number of holographic degrees of freedom
associated with the volume inside the regularized boundary
would be given by:
\be
N_{hdof}=\frac{A(\delta)}{4ln2 G_{5}}
\ee
The quantum field theory ultraviolet cutoff
will now be identified with the gravitational infrared cutoff.
\par
By doing so
we discover, using
Maldacena's expression for the $AdS_{5}$ radius, that the number
of holographic degrees of freedom associated with the volume with 
boundary at $r=1-\delta$ is precisely, up to a numerical factor $\frac{1}{4}$,
the number of degrees of freedom of the regularized $N=4$
supersymmetric Yang Mills theory, with ultraviolet cutoff equal 
$\delta$.

\par

The main physical consequence of this result is to provide
a solid basis for the interpretation
of the holographic coordinate as a renormalization group parameter
\cite{a,ag,pp,gp,dz,bk,fpw,hv,bvv,vv}, thus allowing 
to reinterpret Maldacena's correspondence in terms
of the following
simbolic relation: (Wilsonian QFT Renormalization group in 
$D$ space-time dimensions) =
(Holographic principle in the gravity dual in $D+1$ space-time dimensions).
%%%%%%%%%%%%%%%%%%%%%%%%%%%%%%%%%%%%%%%%%%%%%%%%%%%%%%%%%%%%%%%%%%%%%
\subsection{ Null Geodesic Congruences. }
%%%%%%%%%%%%%%%%%%%%%%%%%%%%%%%%%%%%%%%%%%%%%%%%%%%%%%%%%%%%%%%%%%%%%
Let us consider $AdS_{5}$ in horospheric (conformally flat) coordinates:
\be
ds^ {2}=\frac{R^{2}}{z^{2}}(dz^{2} + dX^{2})
\ee
The boundary is now located at $z=0$. The first thing we need in 
order to check the holographic hypothesis in these coordinates is to define,
in an intrinsic way, a 3-dimensional surface. This can be easily done
using a congruence of null geodesics \cite{ag}. 
In fact given a congruence
of null geodesics, in a space-time of
dimension $D$, with tangent vector $k$, we can define a codimension two
surface as follows. Let $p$ a point on the null geodesic and $T_{p}$
the tangent space at the point $p$. We define $V_{p}$ as the subspace
of vectors in $T_{p}$ which are orthogonal to $k$. Now since $k$ is null 
we define the codimension two quotient space with respect to the equivalence
relation $x-y \sim k$. This space will define the codimension two surface
to be used later in order to define the holographic bound. 
Let us denote $h$ the induced
metric on this surface. It is easy to prove \cite{hawkingellis}
that
\be
\frac{d\sqrt{h}}{d\lambda}=\theta.\sqrt{h}
\ee
where  $\lambda$ is the affine parameter of the null geodesic and
$\theta$ the expansion of the null congruence.
   
\par

Given a null geodesic  $(X(\lambda),z(\lambda))$,
it is easy to 
check that
\be
\sqrt{h}(\lambda)=\frac{R^{3}}{z^{3}(\lambda)}
\ee
Using $z$ as the ultraviolet cutoff for the quantum field theory, we
can the write the holographic hypothesis in more geometrical terms as
\be
\frac{\sqrt{h}}{G_{5}} = \frac{N^{2}}{z^{3 }}
\ee
The general idea of using null geodesic congruences in order
 to define a holographic area
was first introduced in reference \cite{ag}. For a general approach
to holographic entropy based on null congruences see reference \cite{bousso}.
In the appendix we present for completness the explicit construction
of the codimension two volume element in the $AdS_{5}$ case using a different
(although equivalent) approach.
%%%%%%%%%%%%%%%%%%%%%%%%%%%%%%%%%%%%%%%%%%%%%%%%%%%%%%%%%%%%%%%%%
\subsection{Holographic C- function}
%%%%%%%%%%%%%%%%%%%%%%%%%%%%%%%%%%%%%%%%%%%%%%%%%%%%%%%%%%%%%%%%%
Let us consider five dimensional space-time metrics 
preserving four dimensional Poincar\'e
invariance:
\be\label{metric}
ds^2 = a(r)d\vec{x}_{(1,d-1)}^2 + dr^2
\ee
where $d\vec{x}_{(1,d-1)}^2$ is the flat $d$-dimensional Minkowski metric.
\par
Null geodesics are characterized by the null momentum:
\be\label{momento}
k^{\m} =\frac{a_{(0)}}{a} k_{(0)}^{\m}
\ee
where $\m, \n\ldots = 0\ldots d-1$ and
\be\label{momentos}
k^d = (\frac{a_{(0)}}{a})^{1/2}k _{(0)}^d
\ee
and the initial values are indeed null:
\be
a_{(0)}\eta_{\m\n}k_{(0)}^{\m}k_{(0)}^{\n}+ (k_{(0)}^d)^2 = 0
\ee
The expansion
$\theta\equiv \nabla_A k^A$ ( $ A,B \ldots = 0,\ldots d $) is given by:
\be
\theta = \frac{(1-d)k_{(0)}^d a'}{2 a^{3/2}}
\ee
(with $a'\equiv \frac{da}{dr}$),
from which it stems that when $k_{(0)}^d >0$ as well as $a'>0$, then $\theta<0$;
that is, there is convergence of the null congruence.
\par
Ulterior considerations will make use of the tangentially projected 
Ricci tensor:
\be\label{ricci}
R_{AB}k^A k^B = \frac{3 a_{(0)} (k_{(0)}^d)^2 ( (a')^2 - a a'')}{2 a^3}
\ee
The \textbf{null convergence condition}, $R_{AB}k^A k^B\geq 0$ then translates 
into
\be
(a')^2 \geq a a''
\ee
(This is marginally true in the AdS case, which is a maximally symmetric space
so that $R_{AB}k^A k^B = 0$ for a null vector $g_{AB}k^A k^B = 0$) 
\par
In conformally flat coordinates the metric (\ref{metric}) can be writen like
\be
ds^{2}= \Omega^{2}(z)(dX^{2} + dz^{2})
\ee
with $z$ the conformally flat coordinate defined by
\be
dr= \sqrt{a}dz
\ee

Based on our previous discussion we suggest the following definition
of holographic C-function
\be\label{hc}
C= \frac{\sqrt{h}.z^{3}}{G_{5}} = \frac{z^{3} \Omega^{3}(z)}{G_{5}}
\ee
with $\sqrt{h}$ given by
\be
\sqrt{h}= a(r)^{\frac{3}{2}}
\ee
By construction this holographic C-function coincides,
in the $AdS$ conformal case, with the value
(in the large $N$ limit) of the  central extension $N^{2}$ as defined
by the Weyl anomaly.
It is interesting to observe that the invariance of
the C-function in the $AdS$ case is a direct consequence of the scale
invariance of the metric with respect to dilatations: $X\to \lambda X$,
$z\to \lambda z$.

\par

In quantum field theory the C-function \cite{zamo}\cite{cardy} 
is a positive function of the 
couplings $g_{i}$ and on some length scale $L$ that satisfy the 
renormalization group equation, i.e independence on the renormalization
group scale $\mu$

\be\label{rg}
(\mu\frac{\partial}{\partial\mu} + \beta_{i}\frac{\partial}{\partial g_{i}})C(\mu L, g_{i})=0
\ee

The C-theorem is equivalent to the irreversibility of the renormalization group flow,
more precisely

\be\label{L}
L\frac{\partial}{\partial L}C(\mu L, g_{i}) < 0
\ee

Following our approach we will use as our scale $L$ the ultraviolet cutoff 
that we have identified with the conformally flat coordinate $z$. In that way
we easily get the following relation for the geometric C-function
\be
z\frac{\partial}{\partial z}C = 3C ( 1 + \frac{\Omega'(z)}{\Omega(z)}z )
\ee
Remembering  now the value of the null congruence expansion 
$\theta_{AdS} = - \frac{1}{z}$ in AdS
and defining $\theta_{\Omega} = \frac{\Omega'(z)}{\Omega(z)}$ leads to
\be
z\frac{\partial}{\partial z}C = 3C ( 1- \frac{\theta_{\Omega}}{\theta_{AdS}})
\ee
This formula clearly expresses the fact that
 departure from conformal invariance is measured by the ratio
$\frac{\theta_{\Omega}}{\theta_{AdS}}$.
\par
%%%%%%%%%%%%%%%%%%%%%%%%%%%%%%%%%%%%%%%%%%%%%%%%%%%%%%%%%%%%%%%
\subsection{Jacobi Fields}
%%%%%%%%%%%%%%%%%%%%%%%%%%%%%%%%%%%%%%%%%%%%%%%%%%%%%%%%%%%%%%%
It is not difficult to show that any vector field which commutes with the
 tangent vector
to a null geodesic congruence obeys the Jacobi equation
\be
\frac{D^2}{d\lambda^2}Z^A=- R^A_{BCD}Z^C k^B k^D
\ee
The most general \footnote{ Assuming only dependence on the holographic
 coordinate, $r$} field $Z$ such that
\be
[Z,k] =0
\ee
is
\be\label{jacobis}
Z^A =\frac{Z_{(0)}^d}{k_{(0)}^d} k^A + T^A
\ee
where the components of $T$ read:
\be
T^{\m}= Z_{(0)}^{\m} - \frac{Z_{(0)}^d}{k_{(0)}^d} k_{(0)}^{\m}
\ee
and
\be
T^d = 0
\ee 
Jacobi fields form a congruence of their own. Their expansion can be 
fully attributed to the component in the direction of $k$, and is given by:
\be
\theta_J = Z_{(0)}^d a_{(0)}^{1/2}\frac{(d-1) a'}{a^{3/2}}
\ee
It is interesting to notice that the only possible locus of convergence
occurs at $a = \infty$
%%%%%%%%%%%%%%%%%%%%%%%%%%%%%%%%%%%%%%%%%%%%%%%%%%%%%%%%%%%%%%%%%%%%%%%
\subsection{Renormalization Group Flows}
%%%%%%%%%%%%%%%%%%%%%%%%%%%%%%%%%%%%%%%%%%%%%%%%%%%%%%%%%%%%%%%
The simplest example of a renormalization group flow, interpolating
between two conformal field theory fixed points, is described, in the dual
gravity picture, as a kink solution of the five dimensional $SU(4)$ gauged
$N=8$ supergravity (see for instance, the references \cite{gp,dz,fpw}). 
The space-time metric
of a kink solution preserving four dimensional Poincar\'e invariance is
of the type given in equation (\ref{metric}). Based on the 
gravitational description of the conformal anomaly \cite{hs} 
the following C-function was
suggested in references \cite{gp,fpw}
\be\label{c}
C = \frac{cte}{A'^{3}}
\ee
with $e^{2A} = a$
The monotonicity of this function with respect to the coordinate $r$
is guaranteed by
\be\label{condition}
A''< 0
\ee
Now, from the old equation (\ref{ricci}) we see that (\ref{condition}) is
equivalent to the null convergence condition,
namely $R_{a,b}K^{a}K^{b} > 0$ for any null vector $K$. It is interesting
to notice that the appearance of the null convergence condition is already
suggesting the holographic nature of the C-function in the sense 
discussed above. Namely the null convergence condition implies
the focusing of the null congruence of geodesics used to define
the codimension two surface employed to measure the number of holographic 
degrees of freedom. It is important to point out that both
candidates for the C-function, namely the one defined in equation
(\ref{c}) based on the conformal anomaly and the geometric one
defined in equation (\ref{hc}) coincide on the conformal points
and both are monotonic if the null convergence condition is satisfied.
\par
The C-function (\ref{c}) can be written in more geometric terms
as follows
\be
C = \frac{\sqrt{h}}{G_{5} \theta_{\Omega}^{3}}
\ee
The difference with respect to the geometric definition (\ref{hc})
is a difference -from the holographic point of view- in what is playing
the role of ultraviolet quantum field theory cutoff, namely $z$ in one case
or the analog
of a Hubble length $\frac{1}{\theta_{\Omega}}$ in the other.
\par
An alternative geometric description of the C-function (\ref{c}) was given
in reference \cite{sahakian} as
\be
C= \frac{cte}{\sqrt{h}G_{5}|\theta|^{3}}
\ee
 $\theta$ being the null geodesic expansion.

%%%%%%%%%%%%%%%%%%%%%%%%%%%%%%%%%%%%%%%%%%%%%%%%%%%%%%%%%%%%%%%%%%%%%
\subsection{Raychaudhuri's Equation and Conformal Invariance}
%%%%%%%%%%%%%%%%%%%%%%%%%%%%%%%%%%%%%%%%%%%%%%%%%%%%%%%%%%%%%%%%
Let us now consider a box in euclidean $\mathbb{R}^{3}$ of side length equal
 $L$. We are
interested in the behaviour of the volume of the box with respect to 
dilatations $L \to \lambda L$. Obviously we get
\be
\frac{d \lambda^{3} L^{3}  }{d \lambda}  = 3 {\lambda}^{2} L^{3}
\ee
We can define the expansion coefficient $\theta = \frac{3}{\lambda}$. This
expansion trivially satisfies the equation:
\be
\frac{d \theta}{d \lambda} = - \frac{{\theta}^{2}}{3}
\ee
This is Raychaudhuri's equation for the null congruence expansion in
$AdS_{5}$. This simple geometrical fact is the reason for a constant
C- function ( i.e conformal invariance ) in $AdS_{5}$. Namely the 
scaling of null congruences is the same of that of three dimensional
euclidean volume. Any departure of conformal invariance is defined
by adding the effect of ``holographic'' matter fields in Raychaudhuri's 
equation i.e adding the term $- R_{ab}K^{a}K^{b}$. 
\par
Thus from the holographic point of view the variation of the C-function
measures the diffrence in the scaling behaviour of three dimensional
euclidean volume and the three dimensional volume associated with
a congruence of null geodesics.
%%%%%%%%%%%%%%%%%%%%%%%%%%%%%%%%%%%%%%%%%%%%%%%%%%%%%%%%%%%%%%%%%
\section*{Appendix}
%%%%%%%%%%%%%%%%%%%%%%%%%%%%%%%%%%%%%%%%%%%%%%%%%%%%%%%%%%%%%%%

%%%%%%%%%%%%%%%%%%%%%%%%%%%%%%%%%%%%%%%%%%%%%%%%%%%%%%%%%%%%%%%
\subsection*{Conformally Invariant case}
%%%%%%%%%%%%%%%%%%%%%%%%%%%%%%%%%%%%%%%%%%%%%%%%%%%%%%%%%%%%%%
Let us employ the coordinates introduced by
Fefferman and Graham (\cite{fg}),and used in
(\cite{hs}) to reproduce the Weyl anomaly
which enjoy the property that the boundary lies at $\rho=0$:
\be
ds^2 =\frac{1}{\rho}g_{\m\n}(x,\rho)dx^{\m}dx^{\n} + 
\frac{l^2}{4\rho^2}d\rho^2  
\ee
(where $l$ is a length scale associated to the total curvature of 
the spacetime, what we called before the radius, $R$).
We shall denote by $x^A$ the set of the five coordinates,
$x^{\m},\rho$, where greek letters run from $0\ldots 3$ and capital
latin letters from $0\ldots 4$, with $x^4\equiv \rho$.  
We shall furthermore concentrate in the simplest case in which
\be
g_{\m\n}=\eta_{\m\n}.
\ee
It is not difficult to compute the null geodesics in the preceding manifold.
They are given by:
\bea
x^{\m}&=&\frac{l^4 \dot{x}_{(0)}^{\m}\dot{\rho}_{(0)}^2}
{2\dot{x}_{(0)}^2(l^2 \dot{\rho}_{(0)} - 2\dot{x}_{(0)}^2 (\lambda-\lambda_{(0)})}+ 
\nonumber\\
&&x_{(0)}^{\m} - \frac{l^2\dot{\rho}_{(0)} \dot{x}^{\mu}_{(0)}}{2\dot{x}^2_{(0)}} \nonumber\\
\rho&=&\frac{l^6 \dot{\rho}_{(0)}^4}
{4\dot{x}_{(0)}^2(l^2 \dot{\rho}_{(0)} -2 \dot{x}_{(0)}^2
 (\lambda-\lambda_{(0)}))^2}
\eea
There are two different behaviours depending on the sign of  $\dot{\rho}_{(0)}$:
if $\dot{\rho}_{(0)} >o$, the geodesic starts at $(x_{(0)}^{\m},\rho_{(0)})$ at
 $\lambda=\lambda_{(0)}$,and reaches infinity at a finite value of the affine 
parameter $\lambda_c $. These geodesics
 can be thought of as
starting at the boundary $\rho=0$ at $\lambda = -\infty$.
\par
When $\dot{\rho}_{(0)}<0$, instead, the geodesic again starts at
$(x_{(0)}^{\m},\rho_{(0)})$ at $\lambda=\lambda_{(0)}$, but now reaches the
boundary $\mathcal{I}$ at $\lambda = +\infty$.  
They can be thought of as coming
from the infinity at $\lambda = -\lambda_c$.

%\begin{figure}[!ht] 
%\begin{center} 
%  \leavevmode \epsfxsize= 10cm \epsffile{conformal.eps}
%\caption{ Symbolic behaviour of geodesics in $AdS_5$ }
%\label{fig:conforme} 
%\end{center} 
%\end{figure}

\par
The tangent 
vector is given by:
\bea
k^{\m} &=& \dot{x}_{0}^{\m}\rho/\rho_{(0)} \nonumber\\
k^4 &=& \dot{\rho}_{(0)} \rho^{3/2}\rho_{(0)}^{-3/2}
\eea
(where the timelike vector $\dot{x}_{0}^{\m}$ is normalized through 
$\eta_{\m\n}\dot{x}_{0}^{\m}\dot{x}_{0}^{\n} =- \frac{l^2 \dot{\rho}_{(0)}^2}{
4\rho_{(0)}}$).
\par
The optical scalars are easily determined from:
\bea
\nabla_{\m}k_{\n}&=&-\frac{1}{2}\dot{\rho}_{(0)}\eta_{\m\n}(\rho)^{- 1/2}
\rho_{(0)}^{-3/2}\nonumber\\
\nabla_{\m}k_4 &=& \frac{\eta_{\m\n}\dot{x}_{(0)}^{\n}}{2\rho\rho_{(0)}}
\nonumber\\
\nabla_4 k_4&=& \nabla_4 k_{\m}=\frac{\dot{\rho}_{(0)} l^2}{8}(\rho\rho_{(0)})^{-3/2}
\eea
The fact that this tensor is symmetric means that the congruence
is irrotational $\omega = 0$, which in turn conveys the fact that
it is hypersurface orthogonal. The expansion, 
\be
\theta\equiv\nabla_A k^A
= -3/2 \dot{\rho}_{(0)} \rho^{1/2}\rho_{(0)}^{- 3/2}
\ee
This physically means that 
there is gravitational focusing {\em towards} Penrose's 
boundary, $\mathcal{I}$.
\par
What we need now is to determine an area which is naturally associated
to the null congruence. The normal hypersurface suffers from the
ambiguity that the vector $k$ itself belongs to it. Although it is
possible to perform an analysis along these lines (as it is done in
Hawking and Ellis's book), there is a physically more transparent
construction, to which we turn.
\par
We shall introduce a sort of Newman-Penrose {\em f\"unfbein} (except
that, living in an odd unmber of dimensions, complex techniques are
not useful).  For that purpose, we shall first consider another null
vector, $l^A\pd_A$, such that $g_{AB}l^A k^B = -1$, and it could be
propagated in a parallel way along the null geodesic, i.e.,
$k^C\nabla_C l^A =0$.  It turns out that this construction defines a
codimension two hypersurface and besides, that the Lie derivative of
the logarithm of the induced volume element is directly related to the null
congruence expansion $\theta$.
\par
The most general vector $l^A$ obeying the
requeriments as above is given by:
\bea
l^{\m}&=&(l_{(0)}^{\m}-(l^4_{(0)}+\frac{2\rho_{(0)}^2}{l^2\dot{\rho}_{(0)}})
\frac{\dot{x}_{(0)}^{\m}}{\dot{\rho}_{(0)}})(\rho/\rho_{(0)})^{1/2} + 
\nonumber\\
&&
\frac{l_{(0)}^4 \rho \dot{x}_{(0)}^{\m}}{\dot{\rho}_{(0)}\rho_{(0)}}
+\frac{2\dot{x}_{(0)}^{\m}\rho_{(0)}^2  }{l^2 \dot{\rho}_{(0)}^2}
\nonumber\\
l^4&=&l_{(0)}^4 (\rho/\rho_{(0)})^{3/2} - \frac{2\rho_{(0)}
(\rho\rho_{(0)})^{1/2}}{\dot{\rho}_{(0)} l^2}
\eea
where $\eta_{\m\n}k_{(0)}^{\m}l_{(0)}^{\n}= l_{(0)}^4
-c \eta_{\m\n}l_{(0)}^{\m}l_{(0)}^{\n} = 0$
\par
Let us choose the simplest possibility (while keeping all the 
generality in $k^A$), namely 
\be
l^A=\frac{2\rho_{(0)}}{\dot{\rho}_{(0)} l^2}(\frac{1}{2} k_{(0)}^{\m},- 
(\rho_{(0)}\rho)^{1/2})
\ee
It is now a simnple matter to show that the three spacelike vectors
$e_{i}^A$ orthogonal to both $k^A$ and $l^A$ are given by:
\be
e_{(i)}^A=(e_{(i)}^{\m},0)
\ee
where $\eta_{\m\n}e_{(i)}^{\m}e_{(i)}^{\n}=\rho$ and $
\eta_{\m\n}e_{(i)}^{\m}k_{(0)}^{\n}=0$. This in turn yields the $\rho$
dependence , namely $e_{(i)}^{\m}=\rho^{1/2}E_{(i)}^{\m}$. where
$\eta_{\m\n}E_{(i)}^{\m}E_{(i)}^{\n}= 1$ and $
\eta_{\m\n}E_{(i)}^{\m}k_{(0)}^{\n}=0$

\par
The finite equations of the hypersurface 
(which lies on a slice of constant $\rho$)are then given by:
\be
x^{\m} = E^{\m}_{(i)}\xi^i
\ee
Correspondingly, the induced metric is:
\be
ds^2=\frac{1}{\rho}\delta_{ij}d\xi^i d\xi^j
\ee
and the volume element \footnote{
The hypersurfaces are non compact in general, so that their volume is infinite.
We shall mostly be interested in the volume {\em density} instead,
which is  a meaningful (and finite) quantity.
}
scales as
\be
A_{d-2}\sim \sqrt{h}\sim \rho^{- 3/2}
\ee
in such a way that indeed
\bea
\pounds(k) A_{d-2}&\equiv& k^A \nabla_A A_{d-2} \\
&=& 
-\frac{3 \dot{\rho}_{(0)}\rho^{1/2}}{2\rho_{(0)}^{3/2}} A_{d-2}\equiv \theta A_{d-2}\nonumber
\eea
%%%%%%%%%%%%%%%%%%%%%%%%%%%%%%%%%%%%%%%%%%%%%%%%%%%%%%%%%%%%%%%%%
\subsection*{A non conformal case}
%%%%%%%%%%%%%%%%%%%%%%%%%%%%%%%%%%%%%%%%%%%%%%%%%%%%%%%%%%%%%%
The following metric was introduced in \cite{Alvarez}:
\be
ds^2 = \frac{1}{\rho} \eta_{\m\n}dx^{\m}dx^{\n} + \frac{l_c^2}{\rho^4} d\rho^2
\ee
Here $l_c$  sets the scale of the curvature, although in the
present case the latter is of course not constant. Please 
note the striking similarity with AdS in the coordinates just employed;
 the difference stemming
from the power of the holographic coordinate in the last term.
We shall actually employ the coordinate $g\equiv\frac{1}{\rho}$, 
so that
\be
ds^2 = g d\vec{x}_4^2 + l_c^2 dg^2
\ee

\par
The general formulas for  Poincar\'e invariant metrics yield for the tangent
vectors to null geodesics the expression:

\bea
&&k_{\m}=c_{\m}\nonumber\\
&&k_4=\frac{c^0}{\gamma}g^{-1/2}
\eea
with $\eta_{\m\n}c^{\m}c^{\n}= - \frac{(c^0)^2}{\gamma }$ .
corresponding to the parametrized null geodesics:
\bea
&&x^{\m}=x_{(0)}+\frac{2c^{\m}\gamma}{c^0}(\frac{3 c^0}{2\gamma}(\lambda-\lambda_0) + g_{(0)}^{3/2})^{1/3}\nonumber\\
&& g =(\frac{3 c^0}{2\gamma}(\lambda-\lambda_0) + g_{(0)}^{3/2})^{2/3}
\eea

It is plain that, given any point with coordinates $x_{(0)}^{\m},g_{(0)}$,
there is a null geodesic starting at it. This means, in mathematical
terms, that there is a {\em null geodesic congruence}. The vanishing of the
rotation conveys the fact that they are all normal to the codimension one
hypersurface 
\be
c_{\m}x^{\m} + 2\frac{c^0}{\gamma}g^{1/2}= constant.
\ee
Given a curve
\bea
x_{(0)}^{\m} &=& x_{(0)}^{\m}(\lambda)\nonumber\\
g_{(0)}&=&g_{(0)}(\lambda)
\eea
the connecting vector is given by:
\bea
Z^{\m}&=&\frac{\gamma \beta^{\m}}{g}\dot{g_{(0)}}g_{(0)}^{1/2} 
+ \dot{x}_{(0)}^{\m}\nonumber\\
Z^4&=&g_{(0)}^{1/2}\dot{g}_{(0)}g^{- 1/2}
\eea
This vector conmutes with $k^A$, and indeed
\be\label{jacobi}
Z = \frac{\gamma\dot{g}_{(0)}g_{(0)}^{1/2}}{c^0} k + \dot{x}_{(0)}^{\m}\pd_{\m}
\ee
We want now to make explicit the construction of the volume element
of codimension $2$ associated to the null congruence in this case.
The most general null vector $l^A$, normalized in such a way that
\be
k.l = -1
\ee
is
\be
l = l^0 (1,\vec{n} \sin\chi,g\cos\chi)
\ee
where $l^0\equiv (c^0 (1 - \frac{g^{1/2} 
\cos\chi}{\gamma}) - \vec{c}.\vec{n}\sin\chi)^{-1}$. 
The vector $\vec{n}$ is a three-dimensional unit vector,
$\vec{n}^2 = 1$, and $\chi$ is an arbitrary angle.
\par
The spacelike 3-surface orthogonal to both $k$ and $l$ will be given, in terms
of the parameters $g_i$, by:
\be
\frac{\pd\Phi^A}{\pd g^i}g_{AB}k^B = \frac{\pd\Phi^A}{\pd g^i}g_{AB}l^B=0
\ee
Whose general solution is:
\bea
t&=&t_{(0)} + \vec{v}.\vec{n}\sin\chi + g_3 \cos\chi\nonumber\\
\vec{x}&=& - t_{(0)} \vec{n}\sin\chi + \vec{v}\nonumber\\
g&=&= - t_{(0)}\cos\chi + g_3
\eea
where the vector $\vec{v}$ is determined in terms of the parameters $g_i$
through:
\bea
v_1&=&g_1 + [V_{(0)}-t_{(0)}\vec{\b}.\vec{n}\sin\chi -g_3\cos\chi +
\nonumber\\
&&
\frac{2}{\gamma}
\sqrt{g_3 - t_{(0)}\cos\chi}]\frac{n_1\sin\chi-\b_1}
{\sin^2\chi +\b^2-2\vec{n}.\vec{\b}\sin\chi}
\nonumber\\
v_2&=&g_2 + [V_{(0)}-t_{(0)}\vec{\b}.\vec{n}\sin\chi -g_3\cos\chi +
\nonumber\\
&&
\frac{2}{\gamma}
\sqrt{g_3 - t_{(0)}\cos\chi}]\frac{n_2\sin\chi-\b_2}
{\sin^2\chi +\b^2-2\vec{n}.\vec{\b}\sin\chi}\nonumber\\
v_3&=&\frac{g_1(n_1\sin\chi-\b_1)+
g_2(n_2\sin\chi-\b_2)}{\b_3-n_3\sin\chi}\nonumber\\
&&
+[V_{(0)}-t_{(0)}\vec{\b}.\vec{n}\sin\chi -g_3\cos\chi +
\nonumber\\
&&
\frac{2}{\gamma}
\sqrt{g_3 - t_{(0)}\cos\chi}]\nonumber\\
&&\frac{n_3\sin\chi-\b_3}
{\sin^2\chi +\b^2-2\vec{n}.\vec{\b}\sin\chi}
\eea
Not all the vectors $\frac{\pd\Phi^A}{\pd g_i}$ do parallel propagate along
the null geodesics. It is easy to check that  an
arbitrary vector $v_{(0)}^A$ defined at a particular point $g_{(0)}$ 
propagates in a parallel way provided its $g$ dependence is fixed to be:
\be
(v_{(0)}^{\m}\frac{g_{(0)}^{1/2}}{g^{1/2}}+\gamma v_{(0)}^4 \b^{\m}
(\frac{g_{(0)}^{1/2}}{g} - \frac{1}{g^{1/2}}),v_{(0)}^4 \frac{g_{(0)}^{1/2}}
{g^{1/2}})
\ee
This can be shown to imply the necessary condition for the f\"unfbein to be
parallel transported, namely 
\be
\vec{\b}=\vec{n}\sin\chi
\ee
(i.e., only when these conditions are fulfilled, can the f\"unfbein be
parallel transported) which in turn singles out the associated 
spacelike normal 
hypersurface as lying in a slice of constant 
$g$ and, besides
\be
t = \vec{x}.\vec{n}\sin\chi + constant
\ee
The induced metric on the hypersurface is then given simply by:
\be
ds^2 = g(\delta_{ij} - n_i n_j) dx^i dx^j
\ee
in such a way that the volume {\em density} reads
\be
A_{d-2}\sim g^{3/2}\cos\chi 
\ee
 We are now ready to recover in a very explicit way the interpretation
of the {\em expansion} as the logarithmic derivative of the natural
$d-2$ volume element associated to the null congruence:
\be
k^A\nabla_A A_{d-2} =\frac{3 c^0}{2 \gamma}g^{- 3/2} A_{d-2}=
\theta\, A_{d-2}
\ee
%%%%%%%%%%%%%%%%%%%%%%%%%%%%%%%%%%%%%%%%%%%%%%%%%%%%%%%%%%%%%%%%%%%%%%

%%%%%%%%%%%%%%%%%%%%%%%%%%%%%%%%%%%%%%%%%%%%%%%%%%%%%%%%%%%%%%%%%%%%
\section*{Acknowledgments}
This work has been supported by the
European Union TMR programs FMRX-CT96-0012 {\sl Integrability,
  Non-perturbative Effects, and Symmetry in Quantum Field Theory}
and  ERBFMRX-CT96-0090 {\sl 
Beyond the Standard model} as well as
by the Spanish grants AEN96-1655 and AEN96-1664.

%%%%%%%%%%%%%%%%%%%%%%%%%%%%%%%%%%%%%%%%%%%%%%%%%%%%%%%%%%%%%%%%%%%%%%

\end{document}